# Large Transverse Thermopower in Shape-Engineered Tilted Leg Thermopile

Ki Mun Bang[a], Sang J. Park[a], Hyun Yu[a], Hyungyu Jin[a,b,*]

[a]*Department of Mechanical Engineering, Pohang University of Science and Technology (POSTECH), Pohang 37673, South Korea*

[b]*Institute for Convergence Research and Education in Advanced Technology, Yonsei University, Seoul 03722, South Korea*

* Corresponding author.

*E-mail address:* hgjin@postech.ac.kr (H. Jin).

**Abstract**

A transverse thermoelectric (TE) device that employs the Nernst effect can generate an electrical potential by applying a temperature gradient perpendicular to the magnetic field. Anomalous Nernst effect (ANE)-based TE materials have been proposed for transverse TE devices due to their advantage of utilizing internal magnetic field of the materials, which offers the benefit of operating without an external magnetic field. To increase the output voltage of transverse TE devices utilizing ANE, materials with higher ANE coefficients, $S_{ANE}$, are required. Currently, $S_{ANE}$ reaches ~6 μV/K at 300 K; however, it is still lower than the Seebeck coefficient of commercial Bi-Te-based TE materials. As proven in conventional TE research, a meticulous design of device structure has the potential to significantly amplify the output voltage for the given material properties. This study proves the same strategy works for transverse TE devices. We demonstrate that a novel device design, where a shape-engineered tilted-leg thermopile structure is employed, significantly enhance the output voltage in the transverse direction. Owing to shape engineering of the leg geometry, an additional temperature gradient develops along the long direction of the leg, which is perpendicular to the direction of the applied temperature gradient, thereby generating an additional Seebeck voltage $V_{SE}$ that adds to the ANE voltage $V_{ANE}$. We further show that a simple adjustment of electrode position within the device can further increase $V_{SE}$. The tilted leg device with electrode adjustment demonstrates a 990% enhanced transverse output voltage compared to that of conventional rectangular leg thermopile-structured devices, wherein only the ANE occurs. This combined output voltage from both the Seebeck effect and ANE is equivalent to what could be achieved by employing materials with the $S_{ANE}$ value of 22.8 μV/K. This value surpasses the $S_{ANE}$ of state-of-the-art ANE materials and devices currently available. The numerical analysis shows the tendencies of the electrical and thermal outputs of the tilted-leg device, which guides a way to further improve the output voltage. Our study paves a way to develop highly efficient transverse TE devices that can overcome intrinsic materials challenges by utilizing the degree of freedom of device design.

**Nomenclature**

| | |
|---|---|
| RL | Rectangular-leg device |
| RLw-E | Rectangular-leg device with electrode adjustment |
| TLw-E | Tilted-leg device with electrode adjustment |
| $L_0$ | Initial length of the tilted leg (m) |
| $\Delta L$ | Cut off length of the tilted leg (m) |
| $P_{Out}$ | Output power (W) of the thermopile thermoelectric device |
| $Q''$ | Heat flux (W/m$^2$) |
| $R_{CO}$ | Cut off ratio of the tilted leg |
| $S_{SE}$ | Seebeck coefficient (V/K) |
| $S_{ANE}$ | Anomalous Nernst coefficient (V/K) |
| $\Delta T$ | Temperature difference (K) |
| $t$ | Thickness of the tilted leg (m) |
| $V_{Out}$ | Output voltage (V) of the thermopile thermoelectric device |
| $V_{SE}$ | Output voltage (V) generated by Seebeck effect |
| $V_{ANE}$ | Output voltage (V) generated by anomalous Nernst effect |

*Greek symbol*

| | |
|---|---|
| $\theta_{ANE}$ | Anomalous Nernst angle ( = $S_{ANE}/S_{SE}$ ) |

## 1. Introduction

Thermoelectric (TE) devices convert waste heat into electrical energy by exploiting the Seebeck effect, in which a temperature gradient across a material generates a difference in the electric potential. Conventional TE devices that employ the Seebeck effect are constructed with a longitudinal structure, wherein the heat flow and electrical output are parallel to each other [1,2]. N- and p-type TE materials are connected into a Π-type structure that requires a large number of legs to increase the output voltage [3]. The complex structure of a longitudinal TE device inevitably leads to large contact resistances between the multiple TE legs and electrodes, which significantly degrades the device efficiency [4]. In contrast, TE devices that utilize the anomalous Nernst effect (ANE) are built in a transverse structure, wherein an electrical potential difference is induced along the direction perpendicular to those of a magnetic field and a temperature gradient. A transverse TE device can be simply fabricated using a single-leg structure and less electrical contacts than a longitudinal TE device [5,6]. A transverse TE device can be easily integrated on a larger scale using significantly fewer manufacturing steps, resulting in a substantial production cost reduction and a simpler device design [4].

One of the key characteristics of transverse TE devices using ANE is the freedom of selecting various thicknesses for device design, whereas the Seebeck effect requires a sufficient length of material along the temperature gradient to generate a large $\Delta T$ for usable voltage production. As long as a temperature gradient is present, the output voltage is generated perpendicular to both the temperature gradient and magnetic field; therefore, the ANE can produce voltage regardless of the thickness of the material [7–9]. This effect is well suited for use in TE conversion systems made of thin materials, enabling the design of flexible devices that can fit the curved surfaces of a heat source. This advantage makes the transverse TE devices attractive for a wide range of applications, including IoT devices, wearable devices, *etc* [10,11]. Furthermore, the utilization of ANE for power generation in transverse TE devices holds the potential to achieve remarkably high energy conversion efficiency [4]. In Seebeck-based TE devices, the Peltier current is induced in the same direction as the Fourier heat. Thus the Peltier effect works in a way to reduce the overall temperature difference in the TE legs, which degrades the conversion efficiency. In contrast, transverse TE devices utilize the Ettingshausen effect to generate a heat current that opposes the Fourier heat and supports device generation [12], thereby increasing the conversion efficiency of TE devices.

Currently, practical deployment of ANE-based transverse TE devices has been limited by small output voltages, mostly due to small anomalous Nernst coefficients of the state-of-the-art materials, two orders of magnitude lower than the Seebeck coefficients of commercial TE materials. To overcome the intrinsic material challenge in realizing practical transverse TE devices, structural engineering of the device becomes imperative, as the device structure design has been proven in conventional TE research to be an effective way to improve overall device performances [13–16]. Previously, transverse devices utilizing ANE has been reported in a thermopile or coiled structure [5,17–19]. The increase in the material length of the transverse TE device can lead to an increase in output performances. Despite these efforts, the resulting devices continue to exhibit low output voltages, ranging from microvolts to a few millivolts, and low output powers, ranging from picowatts to nanowatts [17,18], which can be attributed to their poor material properties [20–28]. Another approach is by using an artificially tilted multilayer structure based on the Seebeck effect [29–32]. N- and p-type TE materials are alternately stacked in a tilted multilayer structure and can generate a non-uniform temperature distribution along the stacked multilayer structure. A tilted structure is used to change the output voltage perpendicular to the direction of heat flow to generate a transverse output voltage. However, attaining thickness reduction poses a challenge for multilayer TE devices utilizing the Seebeck effect. Moreover, the integration of multilayers imparts elevated complexity to the device compared to the single-leg structure of the ANE device. From the perspective of the ANE device, incorporating the tilted structure into the device utilizing ANE can lead to an enhanced output voltage, while preserving its inherent advantages as a transverse TE device. Therefore, the total transverse output voltage could be further improved by combining the ANE voltage with the Seebeck voltage.

In this study, we propose a novel thermopile TE device by adopting tilted-shaped legs to demonstrate the proof-of-concept of such an idea. We demonstrate that simple geometrical modifications on the thermoelectric legs can lead to a significantly enhanced transverse voltage output, by simultaneously utilizing both Seebeck and anomalous Nernst effects under a unidirectionally applied heat flux. Conventional thermopile devices have rectangular legs and generate an output voltage, $V_{ANE}$, using only an ANE (Fig. 1a). The shape-engineered tilted leg design not only generates a temperature gradient along the direction of applied heat flux, but also induces an additional temperature gradient along the direction perpendicular to that of the applied heat flux. This additional temperature gradient induces a voltage generated by the Seebeck effect $V_{SE}$ which adds to $V_{ANE}$ (Fig. 1b). Electrodes can also be placed at the hot and cold edges of the end faces to generate additional $V_{SE}$ in the transverse direction. By employing such structure engineering, the output voltage can be increased by 990% compared to that of conventional rectangular-leg thermopile devices wherein only ANE occurs. The effective ANE coefficient $S_{ANE}$ calculated from the combined output voltage of the tilted-leg thermopile TE device surpasses that of the state-of-the-art ANE materials and devices [26,33,34]. Moreover, fabricating the tilted leg thermopile consumes less material for the same dimensions compared to the rectangular leg thermopile, thus presenting a cost-efficient solution. We also conducted numerical analysis by changing the material properties and leg dimensions to provide guidelines for the geometric design of the leg, which could facilitate the subsequent implementation of the suggested concept.

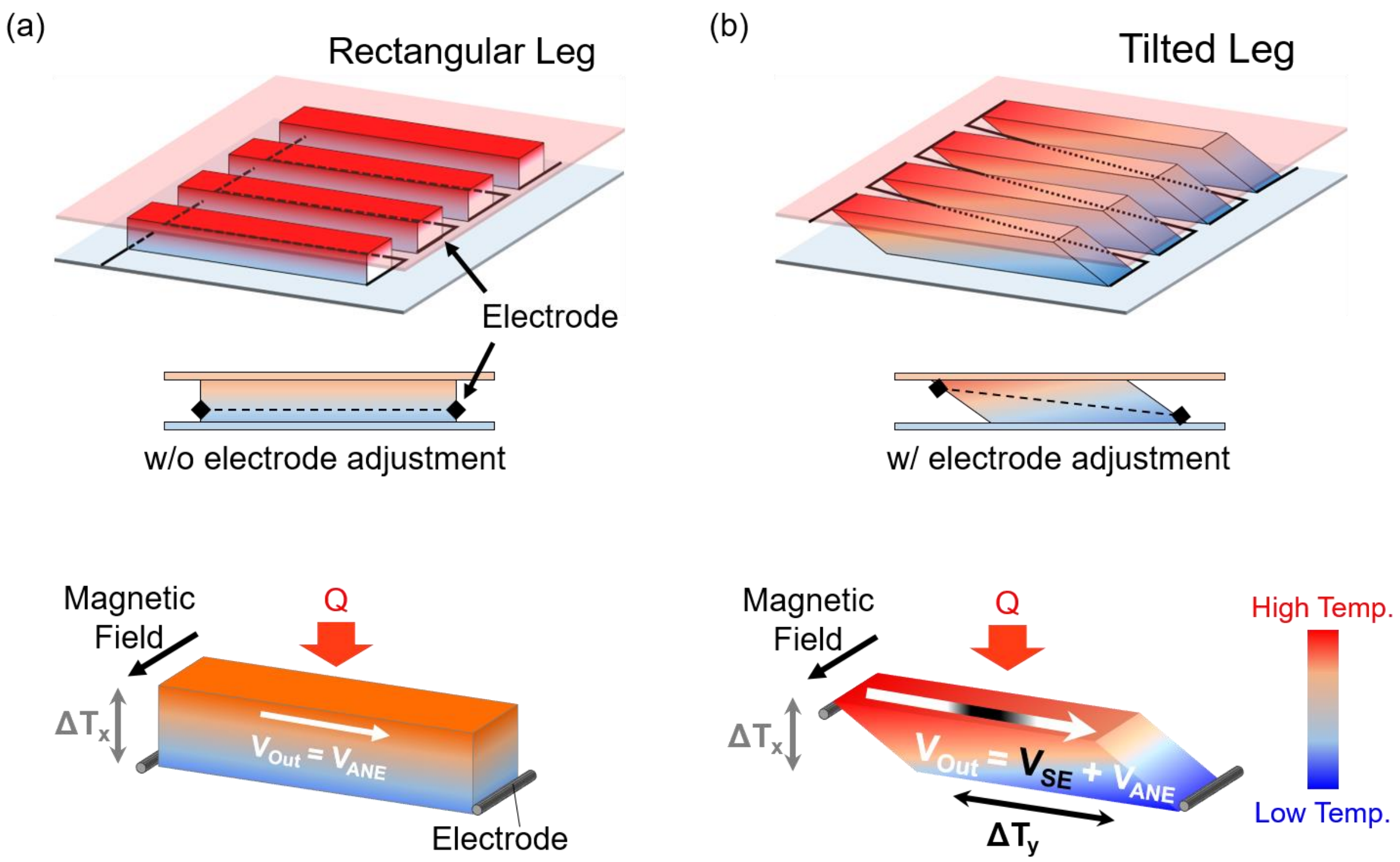

**Fig. 1. Schematic illustration of transverse TE devices that use (a) a conventional thermopile with a rectangular leg without electrode adjustment and (b) a shape-engineered tilted leg with electrode adjustment.**

## 2. Experimental Methods

### *2.1 Fabrication of thermopile device*

Fe powder (Alfa Aesar, ~200 mesh, +99.9 %) and Si powder (Sigma–Aldrich, ~325 mesh, 99 %) were mixed at an 3:1 atomic ratio. The powders were formed into cylinders using a press die at 600 MPa for 10 min. The pressed powders were then melted using an arc melter. The melting process was repeated five times to ensure homogeneity. The sample was sealed in a quartz tube under high vacuum ($<10^{-5}$ Torr) and annealed at 1174 K for 24 h and 973 K for 4 h before water quenching. The sample was cut using a diamond wire saw (MTI Korea, STX-202A) into a 1.8 × 6.66 × 0.88 mm rectangular bar. For the tilted-leg device, the sample was cut diagonally to form a right parallelogrammic prism, where the long sides of the prism represent the parallelograms, hence the term “tilted leg”. The tilted leg was fabricated to fit a rectangle size of 1.8 × 6.66 × 0.88 mm (Fig. 2b, 2c), and four legs were used for each device.

To achieve a uniform temperature distribution, a BeO plate (MTI Korea) was cut into 8 × 10 mm rectangles and utilized as the top and bottom substrates of the heat spreaders. Four legs (both rectangular and tilted) were attached to the bottom substrate using a silver epoxy (Epoxy Technology Inc., EPO-TEK H20E), which was then cured at 400 K for 30 min. Pressure was applied to the top of the legs during curing to uniformly spread the epoxy. The top substrate was then attached to the legs using silver epoxy following the same procedure.

Electrodes were attached to the end faces of the legs. A copper wire (diameter: 0.5 mm) was used as the electrode. The Cu wire was bent into an S-shape, arranged between each pair of legs, attached using silver epoxy, and cured at 400 K for 1 h to complete the fabrication. For the standard rectangular-leg thermoelectric device (RL), the electrodes were attached to the bottom edges of the end faces of the legs (Fig. 2a). For the RL device with electrode adjustment (RLw-E), the electrodes were attached to the bottom of one end face and top of the opposite end face (Fig. 2b). For the tilted-leg device (TLw-E), the electrodes were attached to the acute-angle edges of the end faces (Fig. 2c). The inclusion of RLw-E was implemented to investigate the influence of the top/bottom placement of the electrodes on the TE effects. This configuration generated an additional Seebeck voltage along the rectangular leg, enabling a comprehensive analysis of the TE performance.

### *2.2 Material characterization*

The electrical and thermal transport properties of the $Fe_3Si$ were measured using a customized nitrogen cryostat system (Cryogenics). Two Cu wires (diameter 0.001 in) were attached to the top and bottom of the bar sample for current inputs, and four Cu wires (diameter 0.001 in) were attached to the two edges of opposite sides of the samples for voltage leads. The standard four-probe method was used to measure the electrical resistance. The Constantan wires of a T-type thermocouple (diameter of 0.001 in) were attached to the two edges of one side to estimate the temperature difference along the sample. A 240 Ω resistive heater was attached to the top of the sample for heating. The Seebeck voltage, Nernst voltage, and temperature difference were measured using a nanovoltmeter (Keithley, 2182A), while the magnetic field was continuously swept between ± 12.5 kOe. All measurements were conducted under a high vacuum environment ($<10^{-6}$ Torr). The stoichiometry was characterized using inductively coupled plasma-optical emission spectroscopy (ICP-OES) analysis (AMETEK, Spectro Arcos). The actual composition of $Fe_3Si$ was similar to the nominal design composition. The measured atomic ratios of Fe and Si were 3.02 and 0.99, respectively, with deviations of 0.02 and 0.01 in the atomic ratio. X-ray diffraction (XRD) analysis confirmed the crystal structure of the material to be a DO3 $Fe_3Si$ structure, without any unexpected structural features (Fig. S1).

*2.3 Device characterization*

The voltage outputs $V_{Out}$ of the devices were measured using a home-built He cryostat. To measure $V_{Out}$, each thermopile device was placed on the stage, and a 2000Ω resistance was attached to the device for heating. Thermal grease (Apiezon, H grease) was used to reduce the thermal contact resistance while attaching the device and heater. Currents of 5, 7.07, 8.66, 10, or 11.18 mA were applied to the heater during the measurement, and the experiment was conducted under high vacuum (<$10^{-6}$ Torr) at 300 K. This procedure was repeated for all the devices.

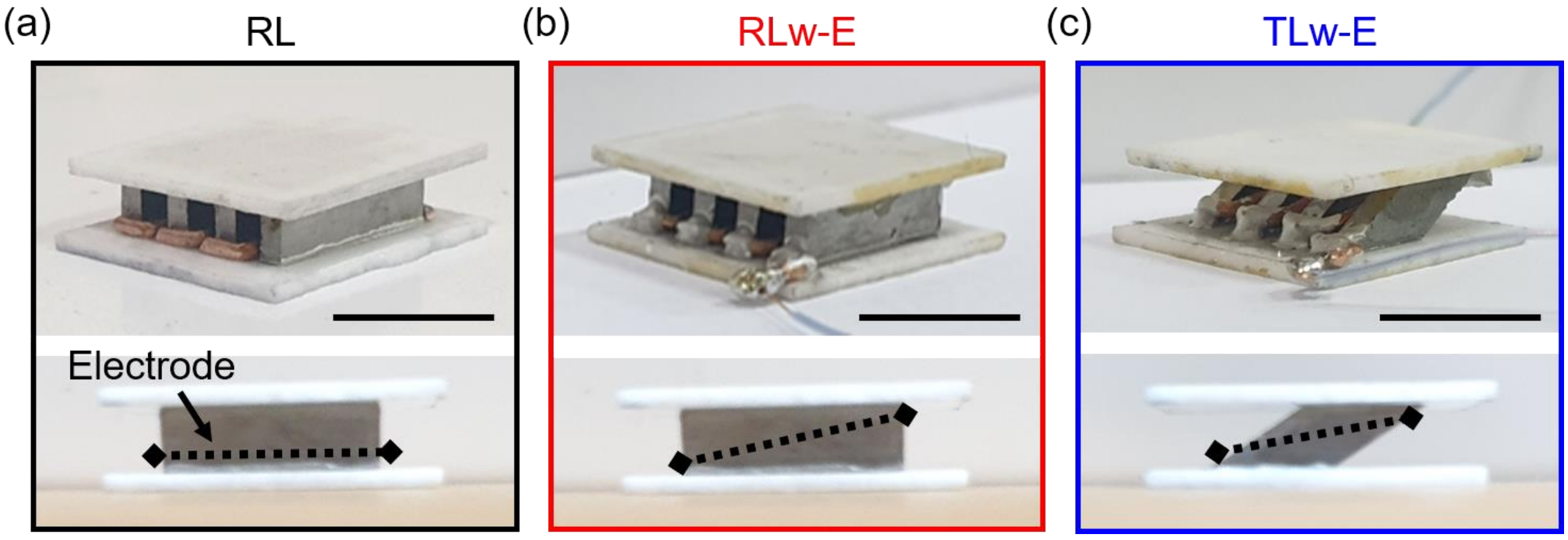

**Fig. 2. Photograph of fabricated 4-pair (a) rectangular-leg device (RL), (b) rectangular leg device with electrode adjustment (RLw-E), and (c) shape-engineered tilted leg device with electrode adjustment (TLw-E). Dotted lines: electrode position. Scale bars: 5 mm.**

## 3. Experimental Results and Discussion

The thermal and electrical properties of the synthesized $Fe_3Si$ samples were measured by varying the temperature from 300 to 400 K (Fig. 3). At 300 K, the $S_{SE}$ was -15.1 μV/K and $S_{ANE}$ was -2.3 μV/K, and the absolute value of both coefficients increased as the temperature increased (Fig. 3a). The thermal conductivity was 32.9 W/m·K at 300 K, which decreased to 31.5 W/m·K at 400 K, while the electrical resistivity was 43.1 μΩ·cm at 300 K and increased to 55.6 μΩ·cm at 400 K (Fig. 3b). $Fe_3Si$ was chosen as the tilted-leg device because of its relatively high $S_{ANE}$ near room temperature [28,35]. Additionally, $Fe_3Si$ is composed of earth-abundant elements [36,37], which can reduce the material costs during fabrication compared to other composite materials with a higher $S_{ANE}$.

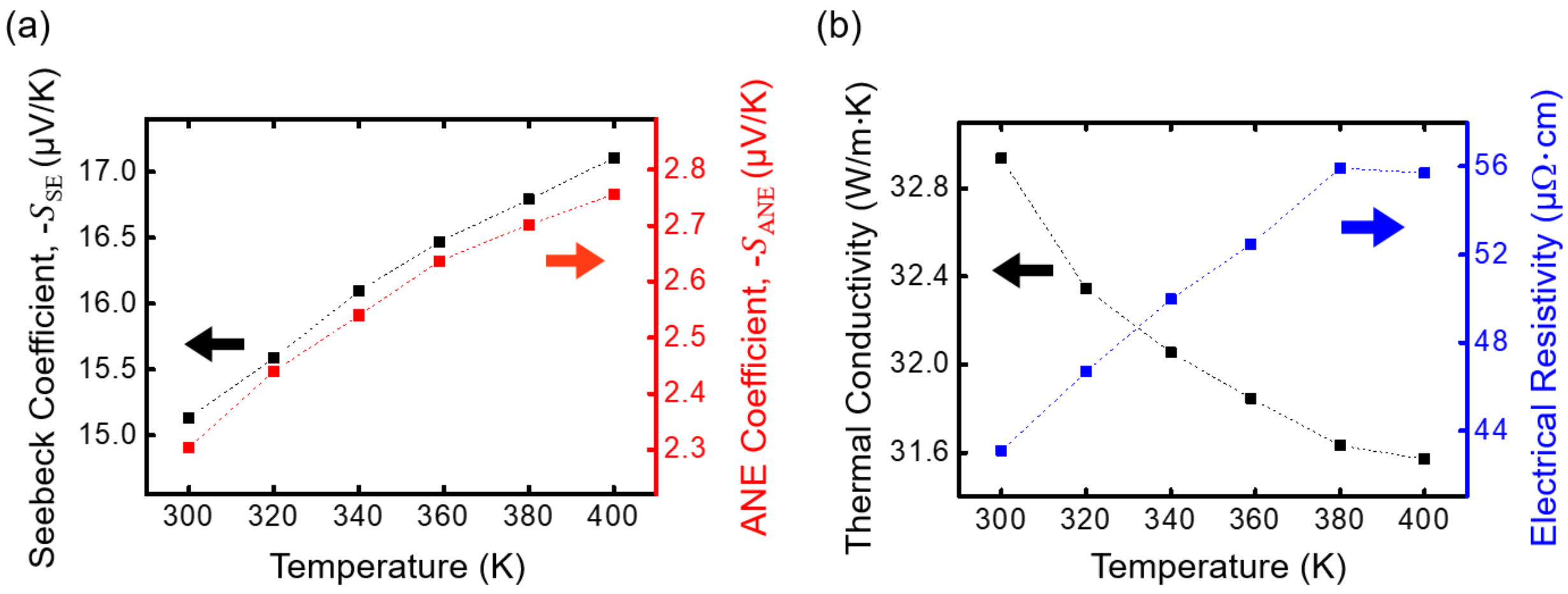

**Fig. 3. (a) Seebeck coefficient -$S_{SE}$, anomalous Nernst coefficient -$S_{ANE}$, (b) thermal conductivity, and electrical resistivity of the $Fe_3Si$ sample.**

$V_{Out}$ was measured for RL, RLw-E, and TLw-E (Fig. 4). Although both Seebeck and ANE coefficients have negative signs, the output voltages are plotted inversely for simplicity. Each $V_{Out}$ value is the sum of $V_{SE}$ and $V_{ANE}$. $V_{SE}$ was generated by the shape of the leg, which shifted the output voltage curve along the Y-axis. $V_{ANE}$ was measured by altering the direction of the magnetic field.

The RL showed $V_{Out}$ = 0.22 ± 1.37 μV, 0.48 ± 2.86 μV, 0.84 ± 4.28 μV, 1.20 ± 5.68 μV, and 1.53 ± 8.12 μV for applied heater powers of 0.03 W, 0.06 W, 0.09 W, 0.12 W, and 0.15 W, respectively (Fig. 4a), and the curves were shifted by 0.22 μV, 0.48 μV, 0.84 μV, 1.20 μV, and 1.53 μV along the Y-axis. In an ideal device, the voltage curve exhibits origin symmetry, indicating that the Seebeck effect does not affect $V_{Out}$. However, the fabricated RL shows a non-uniform temperature distribution owing to the uneven thermal contact resistance resulting from the fabrication process, leading to the generation of a small $V_{SE}$ in the transverse direction.

The RLw-E showed $V_{Out}$ values of 1.53 ± 1.89 μV, 2.66 ± 3.77 μV, 3.99 ± 5.58 μV, 5.17 ± 7.53 μV, and 6.52 ± 10.48 μV for applied heater powers of 0.03 W, 0.06 W, 0.09 W, 0.12 W, and 0.15 W, respectively (Fig. 4b), and the curves were shifted by 1.53 μV, 2.66 μV, 3.99 μV, 5.17 μV, and 6.52 μV along the Y-axis. All these shifts are larger than those in the RL output voltage curves. An additional $V_{SE}$ was generated by changing the location of the electrode diagonally (Fig. 2b).

The TLw-E showed $V_{Out}$ values of 17.0 ± 1.7 μV, 34.5 ± 3.5 μV, 51.3 ± 5.4 μV, 68.8 ± 7.2 μV, and 86.2 ± 9.0 μV for applied heater powers of 0.03 W, 0.06 W, 0.09 W, 0.12 W, and 0.15 W, respectively (Fig. 4c), and each voltage curve was shifted by 17.0 μV, 34.5 μV, 51.3 μV, 68.8 μV, and 86.2 μV along the Y-axis. All these shifts were larger than those in both the curves of RL and RLw-E. The tilted geometry induces localized heat flow and increases $\Delta T$ in the transverse direction, and adjusting the electrode in the diagonal direction generates a $V_{SE}$ together with the $V_{ANE}$. Furthermore, the Seebeck coefficient of Cu is one order of magnitude smaller than that of $Fe_3Si$ [38], which generates a smaller voltage when using $\Delta T$. The Seebeck coefficient of Cu is opposite to that of $Fe_3Si$, and it additionally increases $V_{Out}$ by diagonally placing the electrode for RLw-E and TLw-E.

The maximum, average, and minimum $V_{Out}$ values of each device increased linearly as the heater power increased. The device performance was determined based on its maximum $V_{Out}$. A conventional thermopile device with a rectangular leg only generates $V_{ANE}$ in the transverse direction; therefore, the device performance depends only on the amplitude of the $V_{ANE}$ and cannot be increased by changing the direction of the magnetic field. In contrast, the tilted-leg structure distributes heat nonuniformly along the legs to generate $V_{SE}$, and the saturated voltage is changed by the magnetic field. Therefore, the output voltage can be further increased by selecting an appropriate magnetic field direction.

Compared with the output voltage of RL, the maximum $V_{Out}$ for RLw-E and TLw-E increased to 180% and 990%, respectively (Fig. 4e). This result indicates that using the tilted geometry of the leg can significantly increase the $V_{Out}$ of the device. Using the $S_{ANE}$ of $Fe_3Si$ at 300 K, the effective ANE coefficient $S_{ANE}$ calculated from the combined output voltage is equivalent to that of a device utilizing $|S_{ANE}|$ of 4.05 μV/K and 22.8 μV/K for RLw-E and TLw-E, respectively. The effective ANE coefficient of TLw-E can be achieved through shape engineering of the leg, which is higher than that of state-of-the-art ANE materials at room temperature [27] (Fig. 5). $Fe_3Si$ was selected as the base material in this study because of its earth-abundant composition. The device can further achieve a higher output voltage and equivalent anomalous Nernst thermopower by selecting materials with high $S_{ANE}$ as the base material.

The tilted-leg geometry has the same dimensions as the rectangular leg (Fig. 2) and therefore generates a higher $V_{Out}$ while using less material under the same volume. In addition, RLw-E and TLw-E can operate not only under the magnetic field condition but also in the absence of a magnetic field by exploiting the $V_{SE}$. This versatility allows the devices to be used in a wide range of conditions and applications. The measured internal electrical resistance was 9.68 Ω for RL, 7.41 Ω for RLw-E, and 9.45 Ω for TLw-E. At a heater power of 0.15 W, the maximum output power $P_{Out}$ was 7.7 pW for RL, 34 pW for RLw-E, and 0.95 nW for TLw-E. At a heating power of 1.25 W, TLw-E achieved a maximum $V_{Out}$ of 834 ± 87 μV with a maximum $P_{Out}$ of 89 nW (Fig. S2).

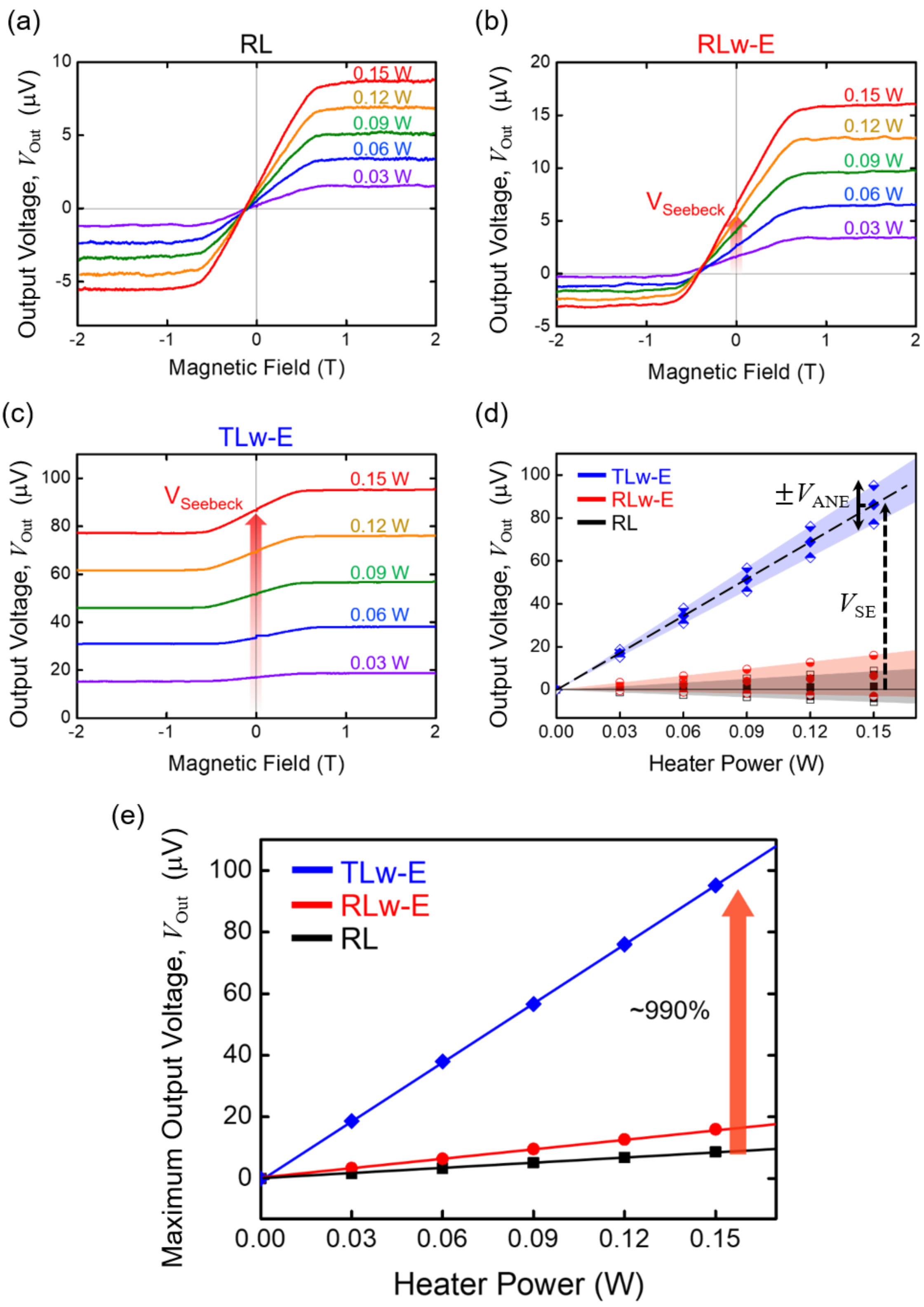

**Fig. 4. Output voltages of the (a) rectangular leg device (RL), (b) rectangular leg device with electrode adjustment (RLw-E), (c) and tilted leg device with electrode adjustment (TLw-E) by changing the magnetic field from -2 T to 2 T. The heater power was applied from 0.03 W to 0.15 W. (d) Ranges of the output voltage for each device vs. heater power**

**(symbol with lower half filled: maximum; filled symbol: midrange; symbol with upper half filled: minimum). (e) Maximum output voltage for each device by changing the heater power. TLw-E shows a 990% increase in the maximum output voltage compared to the RL**

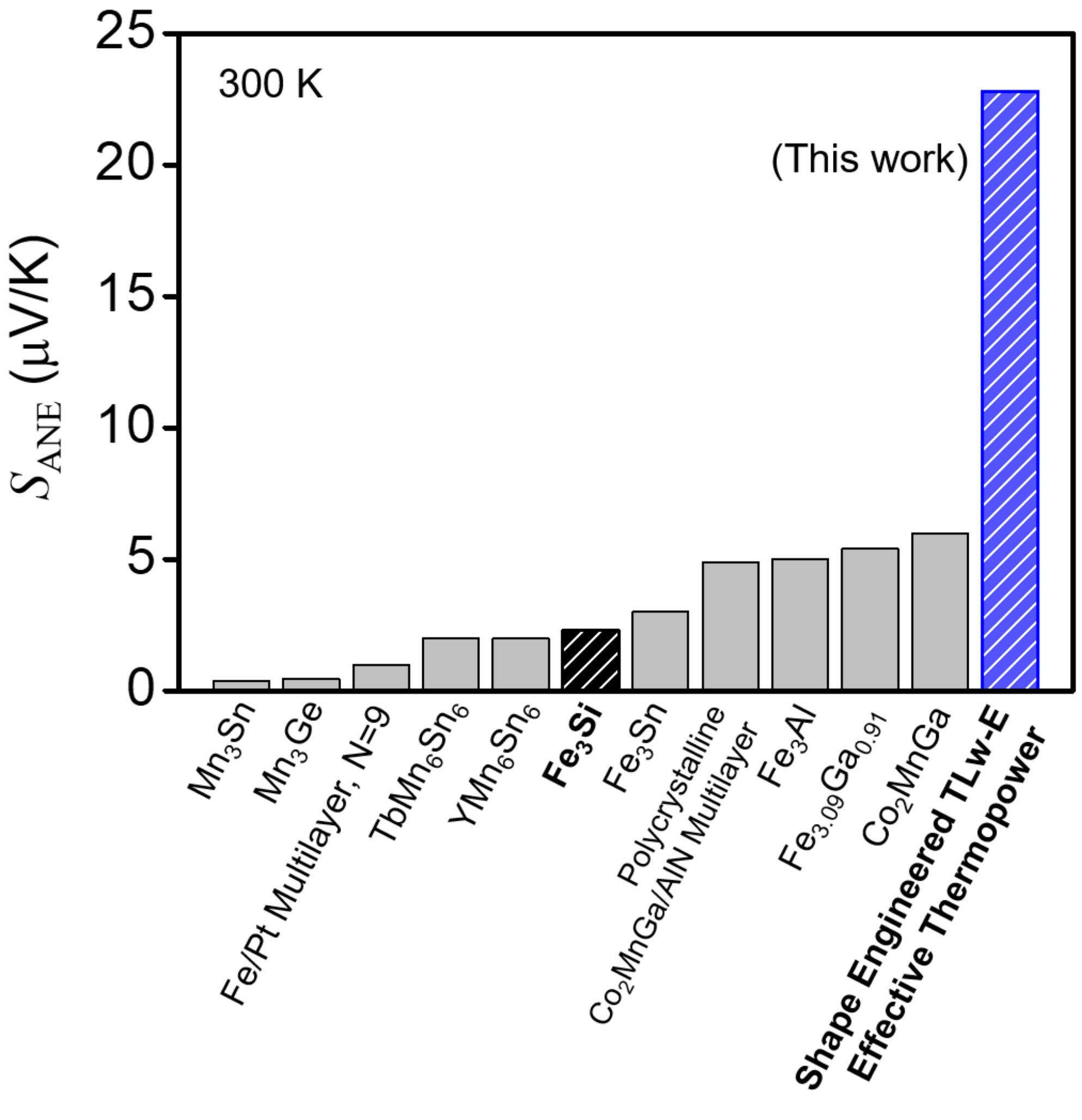

**Fig. 5. Anomalous Nernst thermopower of previously reported materials at room temperature. The data include $Mn_3Sn$** [20]**, $Mn_3Ge$**[39]**, Fe/Pt multilayer** [40]**, $TbMn_6Sn_6$** [41]**, $YMn_6Sn_6$** [42]**, $Fe_3Si$ (Rectangular leg), $Fe_3Sn$** [28]**, polycrystalline $Co_2MnGa$/AlN multilayer** [33]**, $Fe_3Al$** [43]**, $Fe_{3.09}Ga_{0.91}$** [34]**, $Co_2MnGa$** [26]**, and effective thermopower calculated from the combined output voltage of the shape engineered tilted leg device with electrode adjustment (TLw-E) studied in this work.**

## 4. Numerical Analysis

Numerical TE analyses were conducted to evaluate the output efficiencies of thermopile devices with tilted-leg structures and conventional rectangular leg structures. The cut off ratio $R_{CO}$ was used to quantify the tilted geometry of the leg, which is defined as the length $\Delta L$ of the base of the triangular piece cut off as a proportion of the initial length $L_0$ of the leg. During the simulation, the Seebeck coefficient matrix was defined such that the diagonal term represented $S_{SE}$ and the off-diagonal term represented $S_{ANE}$. For simplicity, the absolute values of both $S_{SE}$ and $S_{ANE}$ were used in the matrix. The signs of the coefficients were not considered during the simulation where both $S_{SE}$ and $S_{ANE}$ had negative signs (Fig. 3). The numerical outputs were evaluated by changing $R_{CO}$, thickness $t$ in the x-direction, and $S_{ANE}$ of the leg (Fig. 6a). The location of the electrode was set for two cases: one with electrodes attached to the entire end face (Fig. 6b) and the other with electrodes attached only to the acute edges of the end faces (Fig. 6c).

Simulations were conducted in four steps. First, $V_{Out}$ of the device and the contributions of $V_{SE}$ and $V_{ANE}$ were monitored by changing the anomalous Nernst angle $\theta_{ANE} = S_{ANE}/S_{SE}$ of the system (Section 4.2). Second, the temperature difference was calculated by varying $R_{CO}$ and $L_0$ (Section 4.3). Third, $V_{Out}$, the internal resistance, and $P_{Out}$ were calculated at different locations on the electrode (Section 4.4). Fourth, $V_{Out}$ was calculated for various leg thicknesses (Section 4.5).

### *4.1 Model geometry and materials*

A numerical simulation was conducted for the single-leg device. A rectangular $Fe_3Si$ leg was stacked between the BeO heat spreaders, and a heat flux $Q'' = 100{,}000$ W/m$^2$ was applied to the top surface of the device, while the bottom temperature was fixed at 300 K (Fig. 6a). The thermal and electrical properties of $Fe_3Si$ were obtained from measured outputs (Fig. 3) and references [44], and the material properties of BeO were obtained from the COMSOL material library. $L_0$ was set to 10 mm, and $R_{CO}$ was increased from 0 to 0.5, to form a tilted geometry.

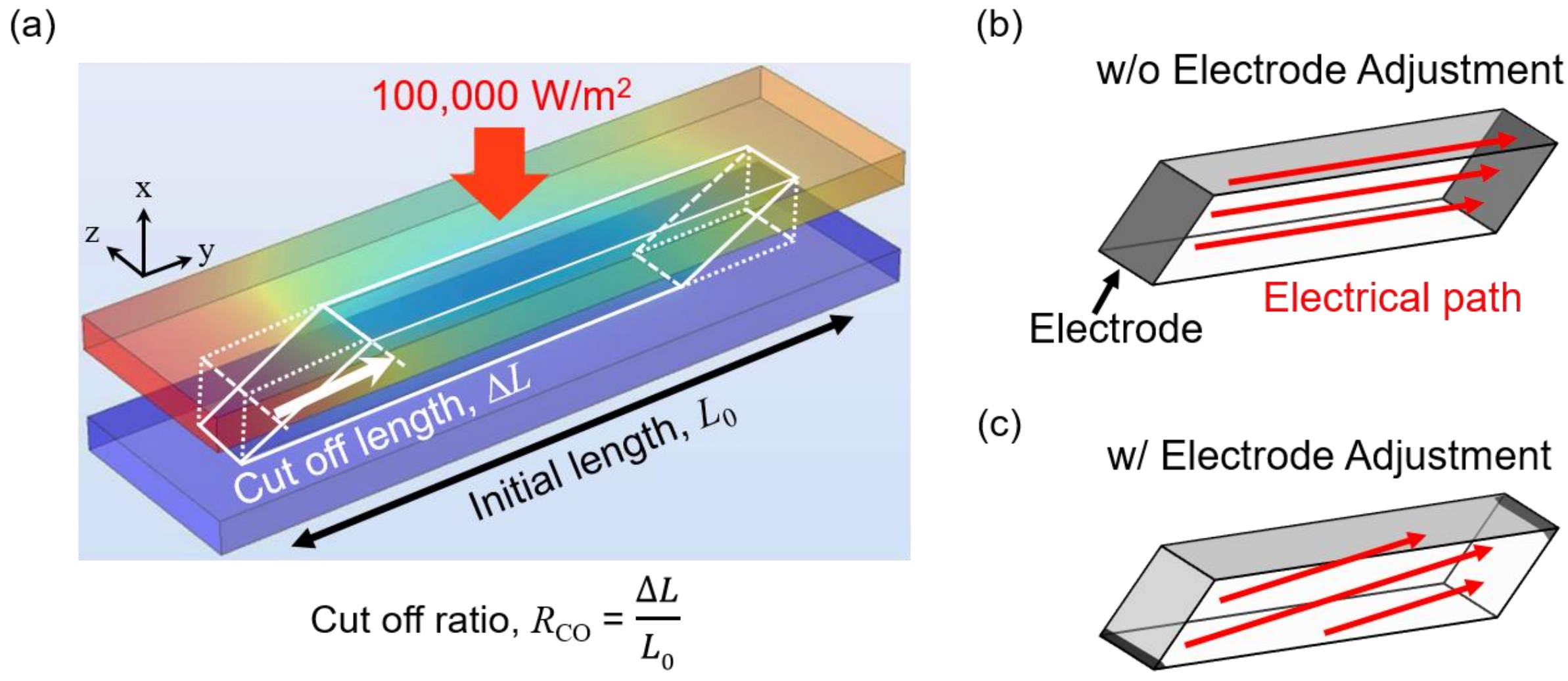

**Fig. 6. (a) Boundary conditions for numerical analysis. Schematic illustration of the (b) leg without electrode adjustment and (c) leg with electrode adjustment.**

*4.2 Evaluation of the output voltage with changes in the cut off ratio and anomalous Nernst angle*

First, to simulate the output voltage and the proportion of the voltage contributed by $V_{SE}$, the anomalous Nernst Angle ($\theta_{ANE} = S_{ANE}/S_{SE}$) was varied (Fig. 7). $S_{SE}$ was fixed at 15 μV/K, and $S_{ANE}$ was varied from 0 to 150 μV/K; that is, $\theta_{ANE}$ was changed from 0 to 10 [45–47]. For all $\theta_{ANE}$, $V_{Out}$ increased as $R_{CO}$ increased (Fig. 7a). $V_{Out}$ increased more rapidly at a low $\theta_{ANE}$ than at a high $\theta_{ANE}$. The proportion of $V_{SE}$ in the total output voltage $V_{Out}$ increased as $R_{CO}$ increased, and its contribution increased rapidly at a low $\theta_{ANE}$ (Fig. 7b).

In addition, $V_{SE}$ and $V_{ANE}$ were simulated with various $R_{CO}$ and $\theta_{ANE}$ by fixing $L_0$ = 10 mm and $S_{SE}$ = 15 μV/K (Fig. 8). $V_{SE}$ increased as the $R_{CO}$ increased but changed negligibly with an increase in $\theta_{ANE}$ (Fig. 8a). The tilted geometry of the leg generates more localized heat at the pointed edges of the leg than at the obtuse edges, which increases $\Delta T$ and leads to the generation of $V_{SE}$. $V_{ANE}$ changed slightly with variations in $R_{CO}$, while it increased as the $\theta_{ANE}$ increased. Owing to the long leg design, the anomalous Nernst effect became more dominant than the Seebeck effect, resulting in an increased proportion of $V_{ANE}$ in $V_{Out}$ (Fig. 8b).

The utilization of the tilted structure can significantly enhance the output performance of current thermopile TE devices for systems with low $\theta_{ANE}$, despite the current TE materials having a lower $S_{ANE}$ than $S_{SE}$, which makes the ANE less ideal as a generator. Therefore, a tilted-leg thermopile device can be used as an intermediate system to connect current conventional thermopile TE devices and materials with large values of $S_{ANE}$.

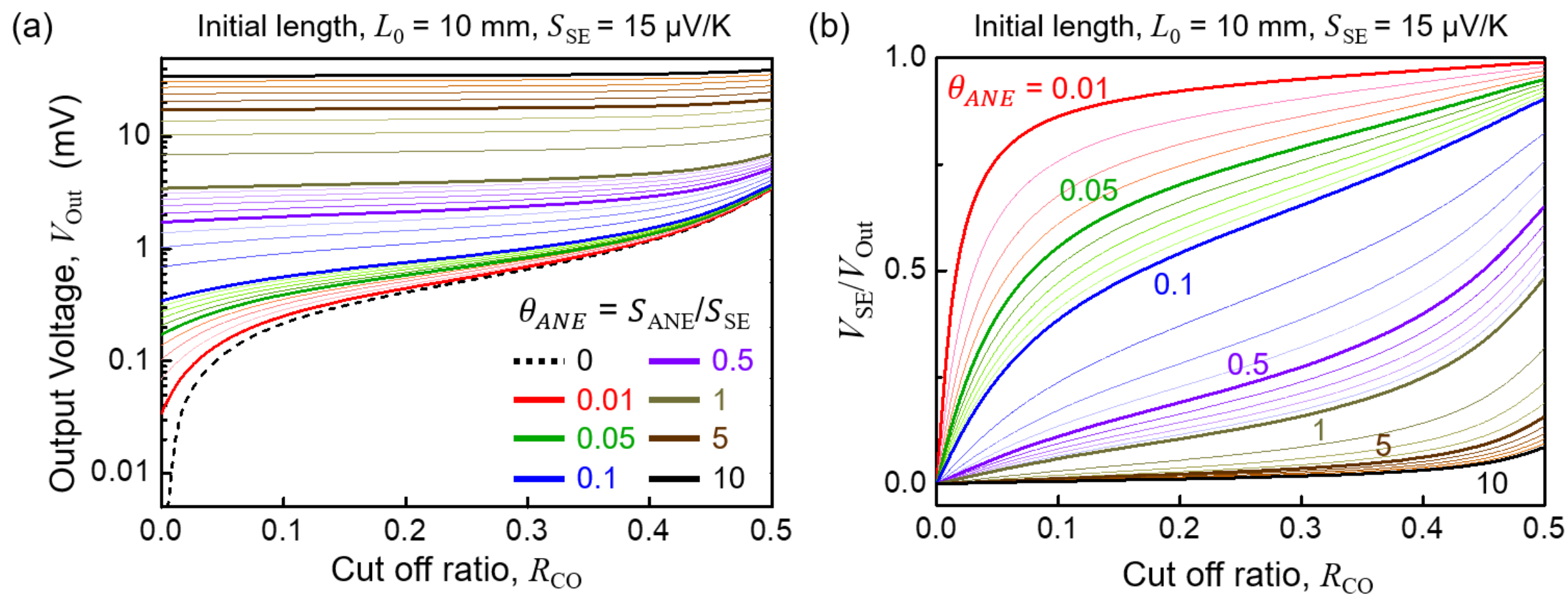

**Fig. 7. (a) Output voltage $V_{Out}$ of the tilted-leg device with change in cut-off ratio $R_{CO}$ and anomalous Nernst angle $\theta_{ANE}$ = $S_{ANE}$/$S_{SE}$. The total length of the device was fixed at 10 mm, and the Seebeck coefficient was fixed at 15 μV/K. (b) $V_{SE}$ as a proportion of total $V_{Out}$.**

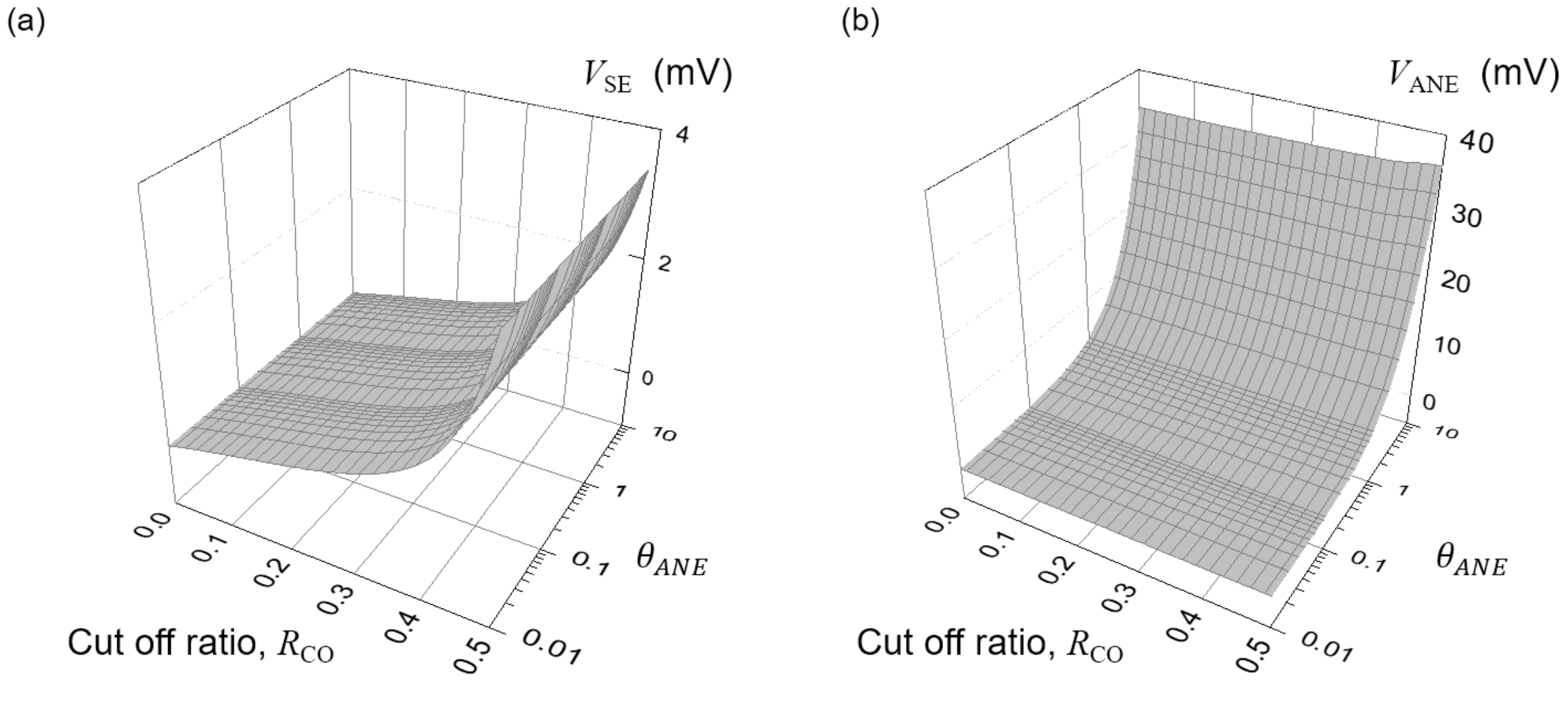

**Fig. 8. (a) Output voltage generated by the Seebeck effect $V_{SE}$ (b) and anomalous Nernst effect $V_{ANE}$ of the tilted leg device with a change in the cut-off ratio $R_{CO}$ and anomalous Nernst angle $\theta_{ANE}$ (= $S_{ANE}/S_{SE}$). The total length of the device was fixed at 10 mm, and the Seebeck coefficient was fixed at 15 μV/K**.

*4.3 Evaluation of the temperature difference with a change in the cut-off ratio and initial length*

$\Delta T$ between the opposite end faces of the leg (Fig.9a)and between the opposite edges of faces (Fig.9b) were calculated for the cases shown in Fig. 6b and 6c,respectively. Two graphs were plotted by changing the $R_{CO}$ and $L_0$ of the device. The average $\Delta T$ for the side area was 0 K when $R_{CO}$ = 0 for all $L_0$, where the leg was rectangular; however, as $R_{CO}$ increased, $\Delta T$ increased to 200 K at $R_{CO}$ = 0.5 (Fig. 9a). The effective area of the thermal heat flow decreased as $L_0$ decreased, and $\Delta T$ increased as $L_0$ decreased.

The $\Delta T$ between diagonally opposite edges of the end faces increased as $R_{CO}$ increased. The device with $L_0$ = 6 mm had the largest $\Delta T$ of 43 K at $R_{CO}$ = 0, owing to the decrease in the effective heat path area in the vertical direction. For all $R_{CO}$ up to 0.5, $\Delta T$ increased as $L_0$ decreased. At $R_{CO}$ = 0.5, $\Delta T$ reached 290 K for all devices. For devices with longer $L_0$, $\Delta T$ increased rapidly as $R_{CO}$ the increased because the tilted-leg geometry became narrower as $L_0$ increased, leading to an increasingly localized temperature distribution. During the device fabrication, a high $\Delta T$ within the leg can degrade the thermal stability of the device. Hence, $R_{CO}$ was limited by the amount of applied heat flux that changed $\Delta T$ along the leg.

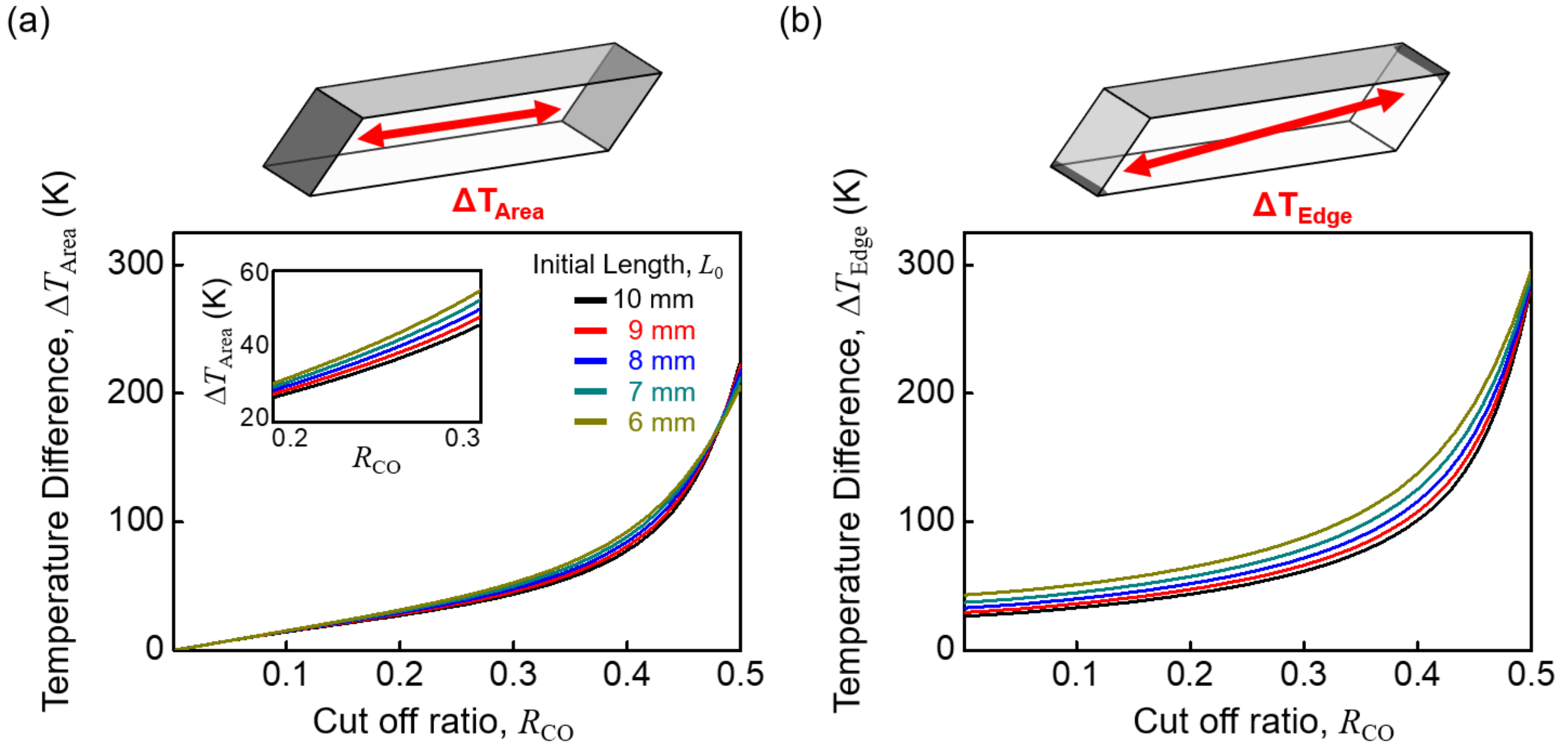

**Fig. 9. (a) Average temperature difference $\Delta T_{Area}$ between the right and left faces of the leg. Inset: $\Delta T_{Area}$ for cut-off ratio $0.2 \leq R_{CO} \leq 0.3$. (b) Temperature difference $\Delta T_{Edge}$ between the opposite edges with changing $R_{CO}$ and initial lengths (line colors) of the leg.**

*4.4 Evaluation of electrical outputs with change of electrode position*

The resistance, output voltage, and output power were collected for legs without electrode adjustment (Fig. 10a, 10c, and 10e) and with electrode adjustment (Fig. 10b, 10d, and 10f). The numerical analyses were conducted using various values of $L_0$ and $R_{CO}$. The electrodes were located on the end faces of the leg (Fig. 6b) and the electrical field passed along the transverse direction. When the electrodes were located at the acute edges of the opposite-end faces, the electrical field passed diagonally between them (Fig. 6c). First, the electrical resistance decreases for the leg without electrode adjustment (Fig. 10a). Then, as $R_{CO}$ increases and a tilted geometry developed, the effective length of the electrical path decreased with a decrease in the resistance. For the leg with electrode adjustment, the electrical path narrowed near the edges of the end face (Fig. 10b). Increasing the $R_{CO}$ yielded a needle-shaped leg that increased the resistance.

In both electrode configurations, $V_{Out}$ increased as $R_{CO}$ increased (Fig. 10c, 10d). At $R_{CO}$ = 0.5, $V_{Out}$ increased to an average of 3.97 mV for the leg without electrode adjustment and to 5.31 mV for the leg with the adjustment. $V_{Out}$ exhibited the same trend as the temperature difference curve (Fig. 9). Therefore, the output power $P_{Out}$ can be calculated for both systems (Fig. 10e, 10f). For the leg without electrode adjustment, an increase in the $R_{CO}$ resulted in the decrease in resistance and an increase in $V_{Out}$. Consequently, this led to a rapid increase in $P_{Out}$ as $R_{CO}$ increased (Fig. 10e). The leg with electrode adjustment achieved a higher $V_{Out}$ than the leg without electrode adjustment, but the increase in resistance reduced $P_{Out}$ as the $R_{CO}$ increased (Fig. 10f).

In the measured results (Section 3), TLw-E exhibits a higher $P_{Out}$ than RL. During the simulation, the electrodes for the tilted leg were attached directly to the acute edges of the end faces, whereas in the fabricated device, the electrode is a bulk system line that is just attached *near* the edges. Therefore, the increase in internal electrical resistance due to the narrowed path will be lower than that in the ideal case, and numerical analysis of the leg with electrode adjustment showed a lower $P_{Out}$ than that of the leg without electrode adjustment.

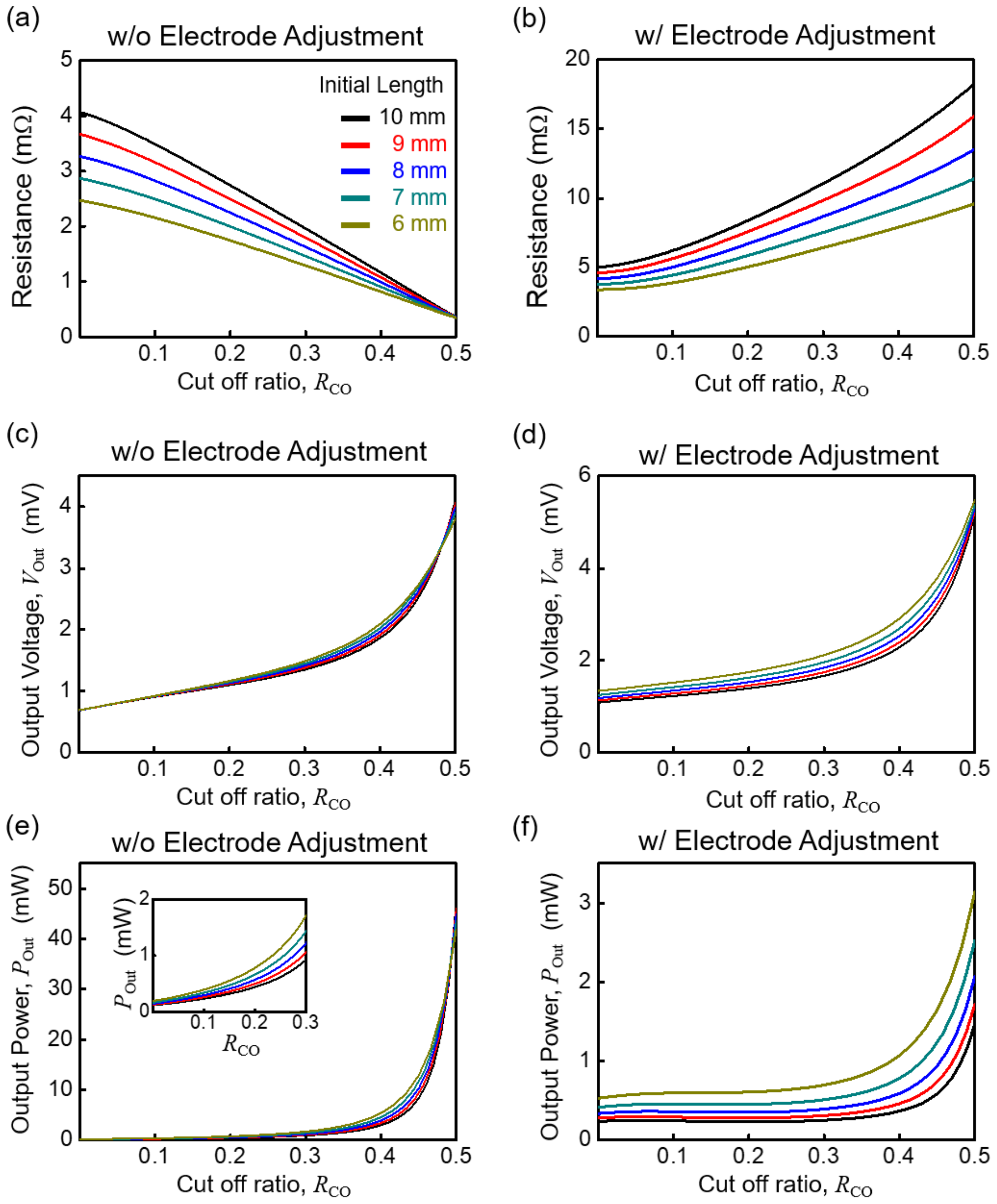

**Fig. 10. Numerical analysis for the resistance, output voltage, and output power (a,c,e) of the device without electrode adjustment, and (b,d,f) the device with electrode adjustment. The total length of the device ranged from 6 mm to 10 mm. Inset: output power for cut-off ratio 0 to 0.3.**

*4.5 Evaluation of output voltage with change of thickness*

To investigate the limit of thickness where the ANE device can generate larger $V_{Out}$ than the transverse Seebeck device, the output voltage was compared between the ANE device and the conventional Seebeck-based multilayered structure transverse device [29–32] while decreasing the thickness from 0.1 mm to 0.001 mm. This analysis aimed to assess and evaluate the effectiveness of the ANE device. The numerical analysis was conducted in 2-dimensions for simplicity (Fig. S3a, S3b). As the thickness decreased, the $V_{Out}$ of the Seebeck device decreased because of the insufficient length across the device to generate a large $\Delta T$ (Fig. 11). However, the ANE device can generate $V_{Out}$ in the transverse direction using only the temperature gradient, and a decrease in thickness does not affect $V_{Out}$. The ANE device exhibited a negligible change in $V_{Out}$ (0.083 mV) over the entire thickness. The intersection where the ANE device outperformed the Seebeck device (Fe/$Bi_2Te_{2.7}Se_{0.3}$ multilayer tilted angle = 50°) occurred near 8 μm. For the ANE device with a thickness from 0.1 mm to 0.001 mm, the $V_{Out}$ increased as the $R_{CO}$ increased (Fig. S3c). Therefore, an increase in $R_{CO}$ can shift the intersection to a higher thickness, where the ANE device outperforms the Seebeck-based transverse device (Fig. 11).

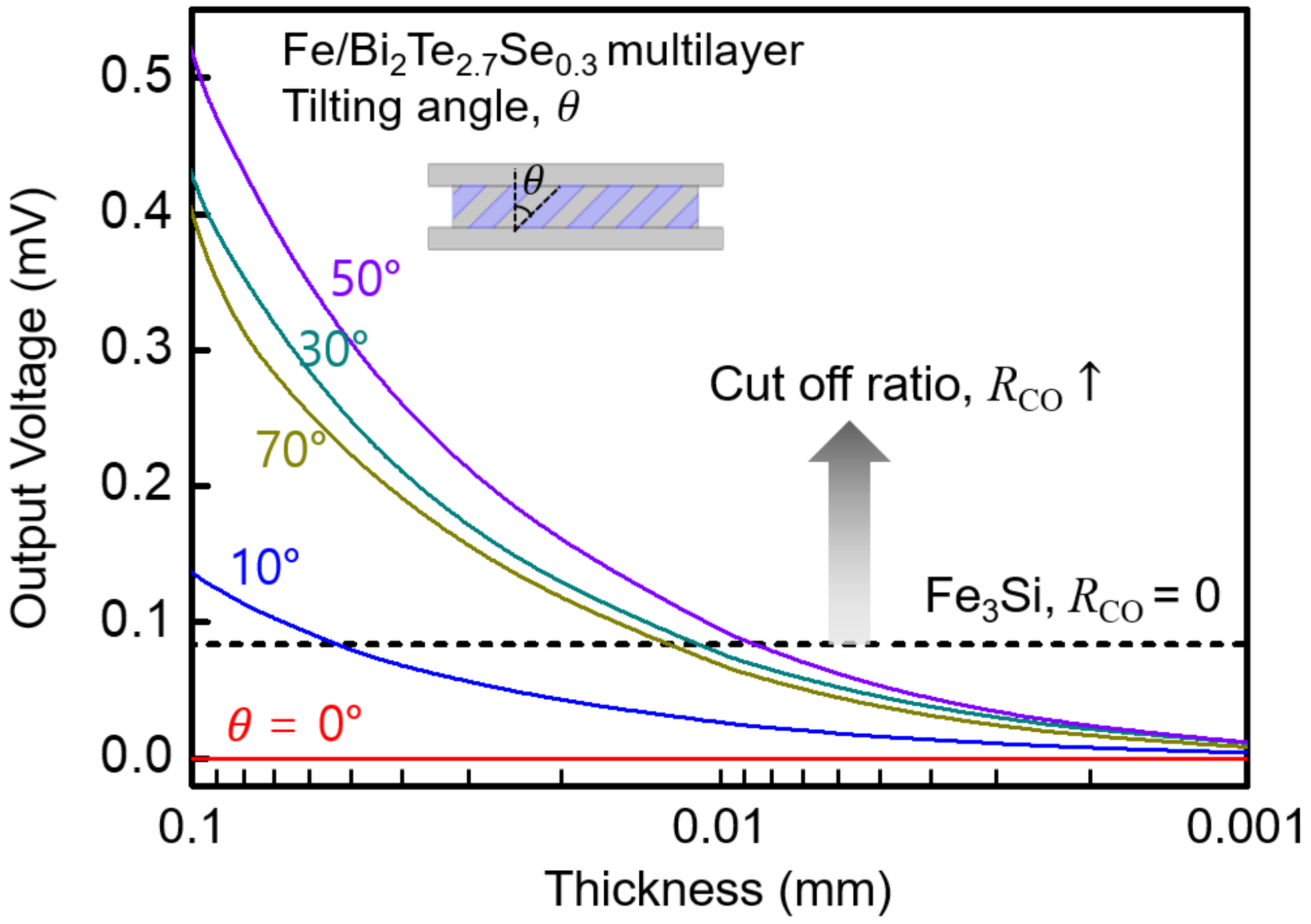

**Fig. 11. Output voltage of the ANE device and the Seebeck-based multilayer tilted device for transverse generation** [31]**. The thickness changed from 0.1 mm to 0.001 mm and the cut-off ratio changed from 0 to 0.5. Fixed values were initial length of device, 10 mm; Seebeck coefficient, 15 $\mu$V/K, and ANE coefficient 3 $\mu$V/K.**

## 5. Conclusions

This study present a shape-engineered tilted-leg thermopile TE generator to increase the output voltage. Compared with a conventional thermopile device with rectangular legs, the geometry of the tilted structure can artificially generate a localized $\Delta T$ along the leg. The shape-engineered tilted leg design not only generates a temperature gradient along the direction of applied heat flux, but also induces an additional temperature gradient along the direction perpendicular to that of the applied heat flux. This additional temperature gradient induces a voltage generated by the Seebeck effect $V_{SE}$ which adds to $V_{ANE}$ in the transverse direction. Additionally, electrodes can be placed at the hottest and coldest edges to generate an additional $V_{SE}$. Compared with the RL, the RLw-E had a 180% higher $V_{Out}$, and the TLw-E had a 990% higher $V_{Out}$, and the effective ANE coefficient $S_{ANE}$ calculated from the combined output voltage is equivalent to that of a device utilizing $|S_{ANE}|$ of 4.05 μV/K and 22.8 μV/K, respectively. The results demonstrate that the use of a tilted geometry of the leg and electrode adjustment can significantly improve the performance of thermopile devices. By generating $V_{SE}$ that is independent of the external magnetic field, the thermopile device can be even used in the absence of a magnetic field. This is in contrast to conventional Nernst-based thermopile devices that typically rely on a magnetic field for their functionality. Moreover, the tilted leg thermopile consumes less material for the same dimension compared to the rectangular leg thermopile, thus offering a cost-efficient solution.

Furthermore, a numerical analysis was conducted to provide geometric guidelines for the leg by changing its material properties and dimensions. A material with a small anomalous Nernst angle $\theta_{ANE}$ can significantly increase $V_{Out}$ by additionally using $V_{SE}$. Currently, the TE material used in ANE devices exhibits a smaller $S_{ANE}$ than conventional devices that utilize Bi-Te. The use of a tilted structure can significantly improve the output performance of a thermopile TE device for systems with small $\theta_{ANE}$. Therefore, a tilted-leg thermopile device can be used as an intermediate system to connect current conventional ANE devices and materials with large values of $S_{ANE}$. Such systems can provide a way to develop highly efficient transverse TE devices for effective energy harvesting across a wide range of applications, including thin and flexible devices.

**CRediT authorship contribution statement**

**Ki Mun Bang**: Conceptualization, Methodology, Validation, Formal analysis, Investigation, Writing - Original Draft, Writing - Review & Editing. **Sang J. Park**: Formal analysis, Investigation, Writing - Review & Editing. **Hyun Yu:** Formal analysis, Investigation, Writing - Review & Editing. **Hyungyu Jin:** Conceptualization, Methodology, Investigation, Resources, Writing - Review & Editing, Supervision, Project administration, Funding acquisition.

**Declaration of Competing Interest**

The authors declare no competing financial interest.

**Acknowledgments**:

This work was partly supported by National Research Foundation of Korea (NRF) grant funded by the Korea government (MSIT) (NRF-2020R1C1C1004291,NRF-2022M3C1A3091988) and by National Research Foundation of Korea (NRF) grant funded by the Ministry of Education, Science, and Technology (NRF-2020K1A4A7A02095438) and by Korea Institute of Energy Technology Evaluation and Planning (KETEP) grant funded by the Korea government (MOTIE) (20188550000290, Development of Meta-Silicide Thermoelectric

Semiconductor and Metrology Standardization Technology of Thermoelectric Power module) and by Basic Science Research Institute Fund (NRF-2021R1A6A1A10042944)

# Large Transverse Thermopower in Shape-Engineered Tilted Leg Thermopile

# Supporting Information

Ki Mun Bang[a], Sang J. Park[a], Hyun Yu[a], Hyungyu Jin[a,b,*]

[a]*Department of Mechanical Engineering, Pohang University of Science and Technology (POSTECH), Pohang 37673, South Korea*

[b]*Institute for Convergence Research and Education in Advanced Technology, Yonsei University, Seoul 03722, South Korea*

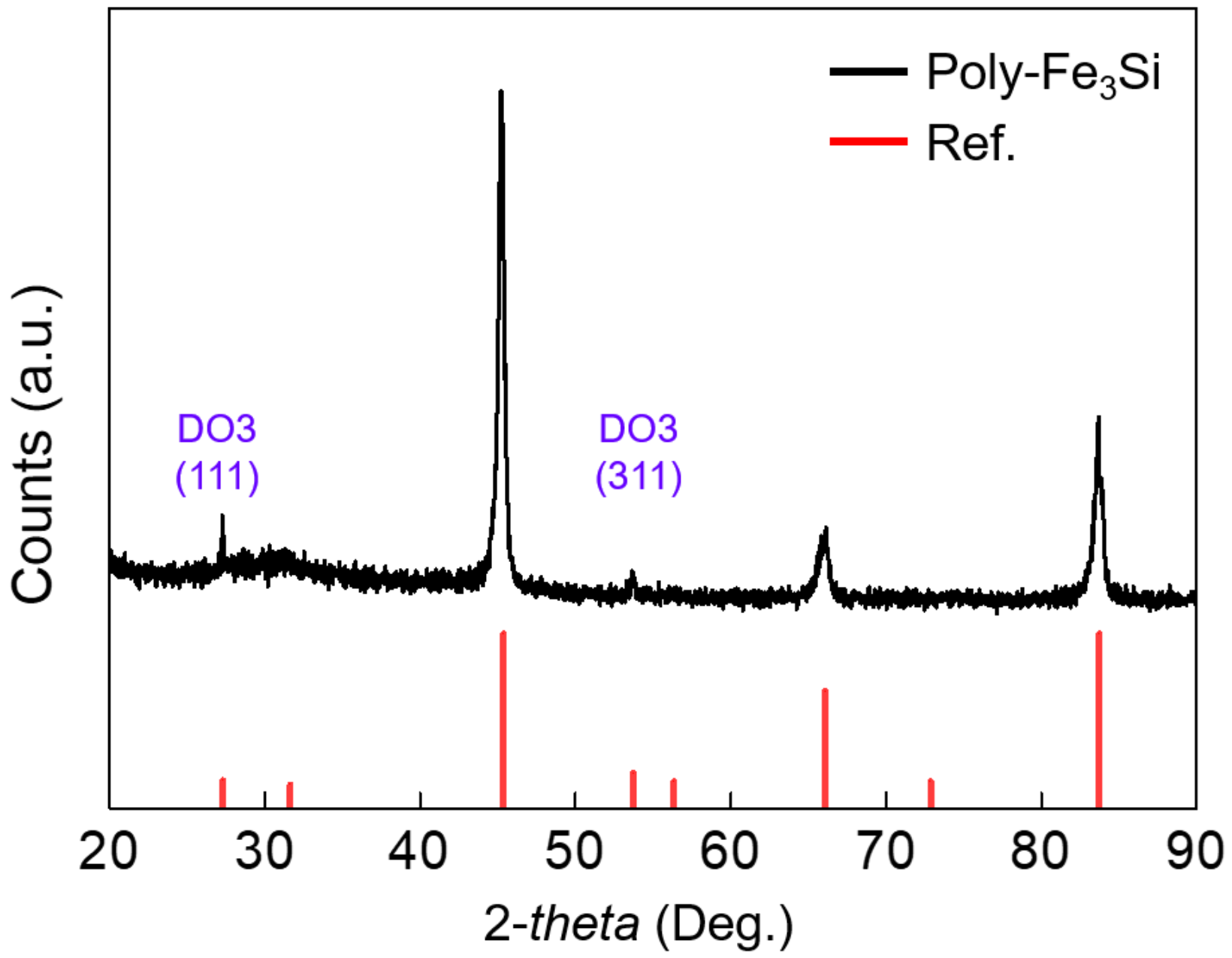

**Fig. S1. XRD pattern of the $Fe_3Si$ sample.**

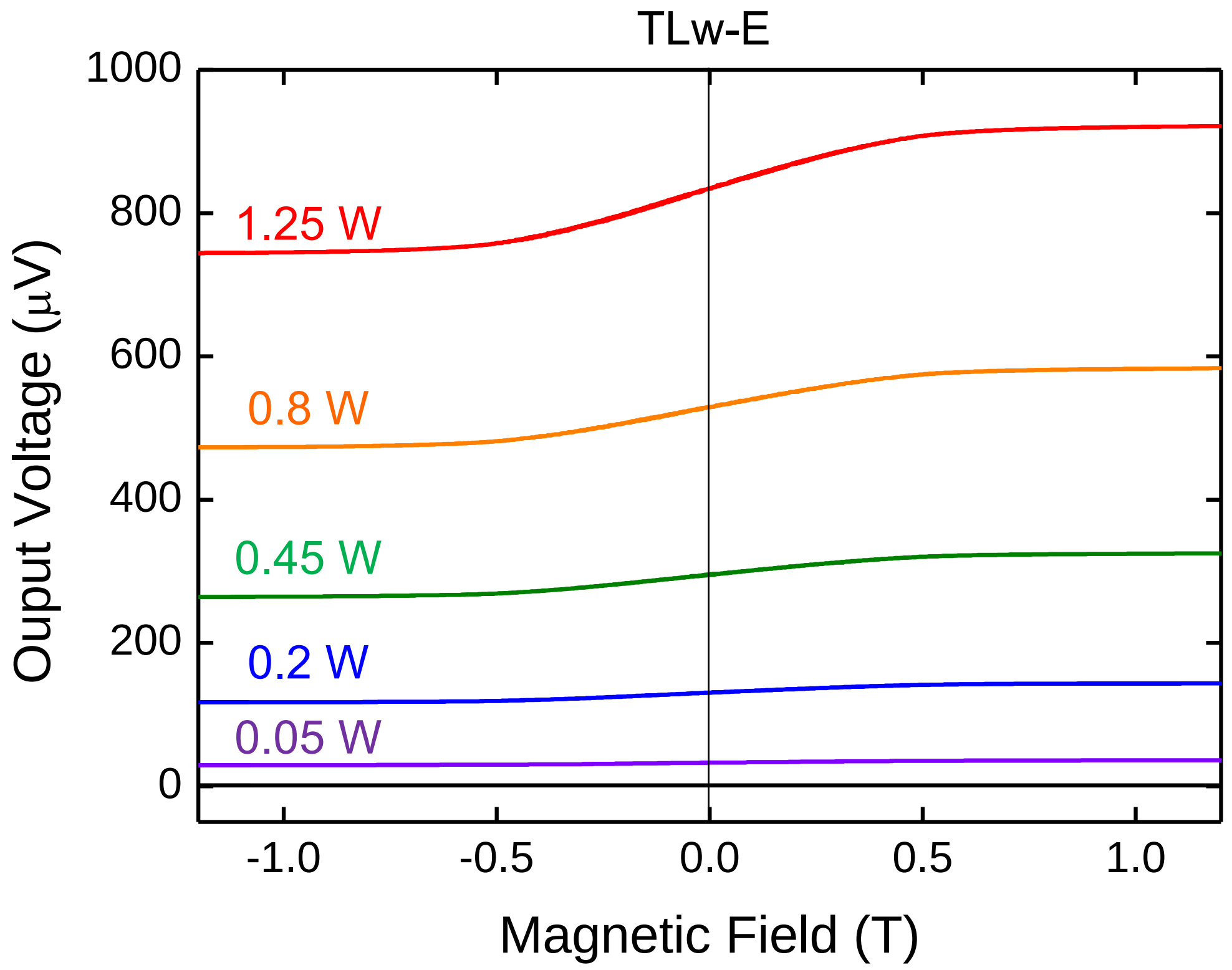

**Fig. S2. Output voltages of the tilted leg device with electrode adjustment (TLw-E) by changing the magnetic field from -2 T to 2 T. The heater power was applied from 0.05 W to 1.25 W.**

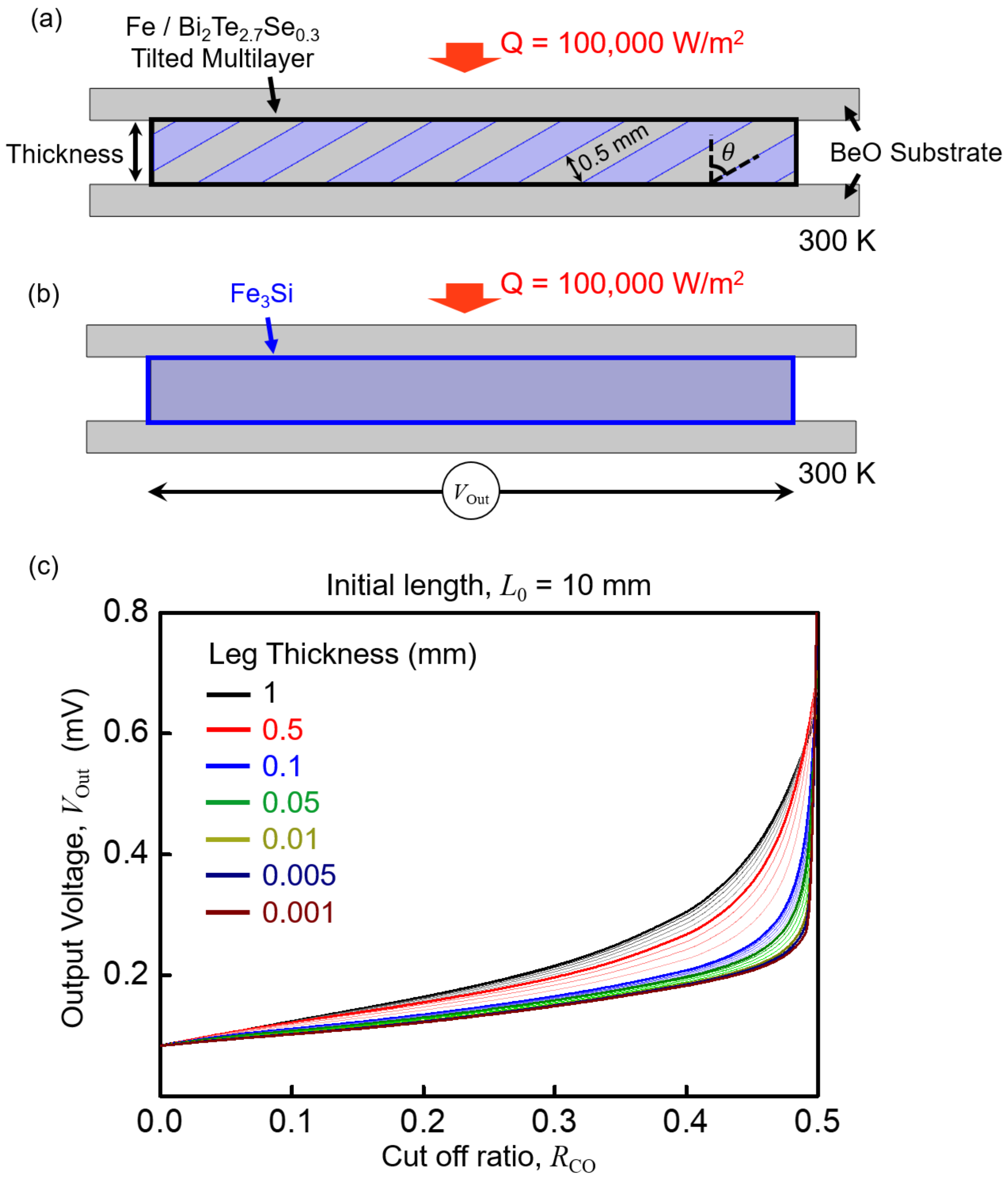

**Fig. S3. Boundary conditions of (a) the Seebeck-based multilayer tilted device and (b) the ANE device. (c) Output voltage of the tilted-leg device by changing the thickness from 1 mm to 0.001 mm and cut-off ratio from 0 to 0.5. The fixed values were the initial length of the device, 10 mm; Seebeck coefficient, 15 $\mu$V/K; and ANE coefficient, 3 $\mu$V/K.**

The $V_{Out}$ of the device was analyzed by changing its thickness and $R_{CO}$ while keeping $L_0$ fixed at 10 mm. The $V_{Out}$ increased rapidly with increasing $R_{CO}$. The ANE device uses a thermal gradient for voltage generation, which is independent of the device thickness, and $V_{Out}$ showed a negligible change as the thickness decreased for $R_{CO}$ = 0.